%% file: hotnets24-nada.tex
\documentclass[sigconf,10pt,nonacm]{acmart}
\newcommand{\para}[1]{\smallskip\noindent\textbf{#1}~}

\usepackage{cuted}
\usepackage{multicol}
\usepackage{lipsum}
\usepackage[many]{tcolorbox}
\usepackage{subcaption}
\usepackage{graphicx}
\usepackage{enumitem}
\usepackage{booktabs}
\usepackage{xspace}

\newtcolorbox{prompt}[2][]{
  title=#2,
  colback=black!1,
  colframe=black!80,
  breakable,
  left=3pt,
  right=3pt,
  top=1pt,
  bottom=1pt,
  boxrule=0.3mm,
  #1,
}

\colorlet{best}{green!20}
\colorlet{worst}{red!20}
\colorlet{good}{green!40!black}
\colorlet{bad}{red!60!black}

\newcommand{\sysname}{\textsc{Nada}\xspace}

\title{Designing Network Algorithms via Large Language Models}

\author{\large Zhiyuan He$^1$, Aashish Gottipati$^{2}$, Lili Qiu$^{12}$, Xufang Luo$^1$, Kenuo Xu$^{3}$, Yuqing Yang$^1$, Francis Y. Yan$^{14}$}

\affiliation{
  \vspace{5pt}
  \institution{$^1$Microsoft Research, $^2$UT Austin, $^3$Peking University, $^4$UIUC}
  \country{}
  \vspace{15pt}
}

\thanks{$^*$Lili Qiu and Francis Y. Yan are the corresponding authors.}
\thanks{$^*$Aashish Gottipati and Kenuo Xu contributed to this work during their internships at Microsoft Research.}

\makeatletter
\renewcommand{\@titlefont}{\LARGE\sffamily\bfseries}
\makeatother
\renewcommand{\authors}{Zhiyuan He, Aashish Gottipati, Lili Qiu, Xufang Luo, Kenuo Xu, Yuqing Yang, and Francis Y. Yan}
\renewcommand{\shortauthors}{Z. He, A. Gottipati, L. Qiu, X. Luo, K. Xu, Y. Yang, F. Y. Yan}

\begin{document}

\input{sections/abstract}

\maketitle

\input{sections/introduction}
\input{sections/approach}
\input{sections/evaluation}

\input{sections/insight}
\input{sections/discussion}
\input{sections/conclusion}

\bibliographystyle{ACM-Reference-Format} 
\bibliography{hotnets24-nada}

\end{document}

%% file: sections/abstract.tex
\begin{abstract}
We introduce \sysname, the first framework to autonomously design network
algorithms by leveraging the generative capabilities of large language models (LLMs).
Starting with an existing algorithm implementation,
\sysname enables LLMs to create a wide variety of alternative designs
in the form of code blocks.
It then efficiently identifies the top-performing designs through
a series of filtering techniques,
minimizing the need for full-scale evaluations and significantly reducing
computational costs.
Using adaptive bitrate (ABR) streaming as a case study,
we demonstrate that \sysname produces novel ABR algorithms---previously
unknown to human developers---that consistently outperform the original algorithm
in diverse network environments, including broadband, satellite, 4G, and 5G.
\end{abstract}

%% file: sections/introduction.tex
\section{Introduction}

Network control and adaptation algorithms have traditionally relied on
human-designed heuristics or, more recently, reinforcement learning (RL).
Notable examples of these algorithms include adaptive bitrate (ABR)
streaming~\cite{yin2015control, mao2017neural,puffer},
congestion control (CC)~\cite{jay2019deep,pantheon},
and load balancing~\cite{mao2019park, xia2022genet}.

As network technology rapidly evolves, there is a growing need
for tailoring network algorithms to specific environments.
For instance, ABR was originally designed for 3G and broadband
networks~\cite{yin2015control, mao2017neural}, but the advent of
more dynamic 4G and 5G networks has prompted the development of
novel, specialized ABR
algorithms~\cite{kumar2016adaptive,tuysuz2020qoe,tran2020bitrate,ramadan2021case}.
Similarly, the emergence of Low Earth Orbit (LEO) satellite
networks has further spurred
customized algorithms~\cite{zhao2023qoe}.

However, developing new algorithms for constantly evolving 
network environments demands substantial expertise and effort.
Motivated by the impressive generative power of large language models
(LLMs),
we explore the following question: \textit{Can we leverage LLMs to automate the design of novel network algorithms tailored to diverse environments?}
We propose that LLMs have the potential to dramatically accelerate
innovation in network algorithm design.

\begin{figure*}[t]
    \centering
    \includegraphics[width=0.95\linewidth]{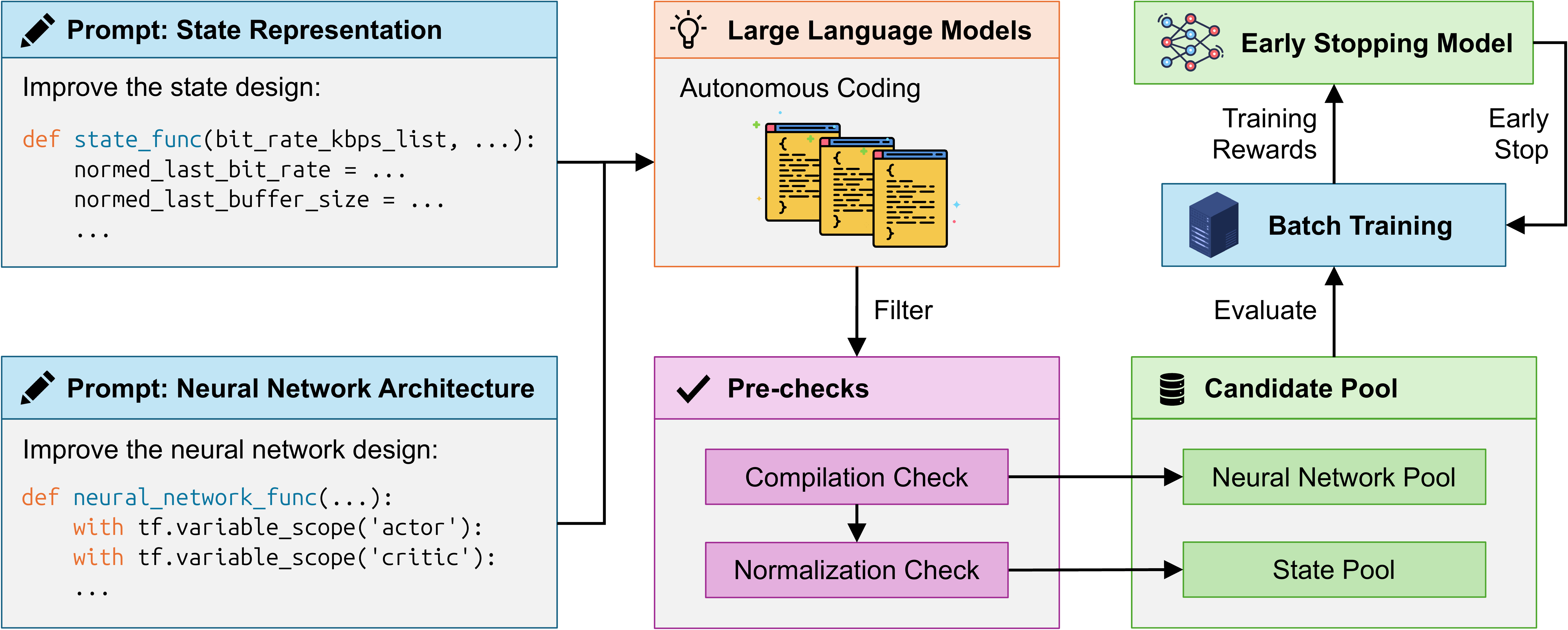}
    \vspace{-2pt}
    \caption{\sysname workflow. It leverages LLMs to generate a wide
    range of alternative designs for a network algorithm and employs a series
    of filtering techniques to efficiently select the most promising designs
    for further evaluation.}
    \label{fig:architecture}
\end{figure*}

The most straightforward approach to utilizing LLMs is by prompting
them to generate new algorithm designs in natural language.
However, after considerable experimentation, we find it challenging
to have LLMs produce high-quality algorithm descriptions
for a target environment (e.g., 5G or satellite networks).
Although LLMs possess general knowledge about these networks,
their responses (even with pseudocode) are often too broad and
lack necessary details,
making it difficult to validate the proposed ideas.

Rather than relying on LLMs to generate algorithm descriptions,
we turn to their remarkable code generation capabilities.
Recent studies have shown that LLMs are proficient at producing
code from human
instructions~\cite{zhou2022docprompting,chen2021evaluating,hong2023metagpt}.
Nevertheless, the code they generate may fail to compile or execute,
contain design flaws, or perform poorly in practice.
Without efficient mechanisms to evaluate the quality of LLM-generated
algorithms (code implementations), the cost of testing would be prohibitively expensive.

We present \sysname (Network Algorithm Design Automation via LLMs),
a generic framework aimed at automating the development of novel
network algorithms using LLMs.
\sysname is applicable to any network algorithm that satisfies two basic 
conditions: it has a functional code implementation,
and its performance can be measured through a network simulator or emulator.
A broad range of network algorithms satisfy these requirements,
and in this paper, we use ABR algorithms in video streaming as a case study.

A widely tested ABR algorithm that meets the above criteria is 
Pensieve~\cite{mao2017neural}. It is based on deep RL,
with an architecture illustrated in Figure~\ref{fig:pensieve}.
Before applying \sysname, we first identify two key components in Pensieve's
design---the RL state representation and the neural network architecture.
Starting from the existing functions (code blocks) that implement these
components, \sysname instructs and stimulates LLMs to generate diverse design 
alternatives (also in the form of code blocks), using carefully crafted prompting
strategies (\S\ref{ssec:sample}).
Next, to efficiently and accurately evaluate a large volume of LLM-generated
designs without incurring excessive computational costs,
\sysname employs a series of filtering techniques, including a compilation check, a normalization check, and an early-stopping
mechanism, to proactively terminate the evaluation of unpromising designs
(\S\ref{ssec:filter}).
Only the remaining designs---a small subset of the total---are then evaluated
at full scale (i.e., trained until convergence). This approach
substantially lowers the overall computational costs.

To assess the effectiveness of \sysname, we gather real-world traces
from various network environments, including broadband, satellite, 4G, and
5G. In each scenario, we find that \sysname is able to generate ABR algorithms
that outperform Pensieve's original design (\S\ref{sec:eval}).
Some of these LLM-generated algorithms offer novel insights into
the design of RL-based ABR algorithms, particularly with regard to
normalization strategies and feature engineering for RL states
(\S\ref{sec:insight}).

In summary, this paper outlines the process of using LLMs to design
novel network algorithms (Figure~\ref{fig:architecture}).
The proposed framework, \sysname, solicits a wide array of alternative designs
from LLMs based on an existing algorithm,
and then employs filtering techniques to efficiently
evaluate their performance and identify the most promising designs.
Using ABR as a case study, we showcase the potential of \sysname
to create network algorithms that outperform existing solutions.
Moving forward, we plan to extend \sysname to other network algorithms,
such as congestion control~\cite{jay2019deep}, and explore its applicability
to non-RL methods.
We hope this work paves the way for further research and ultimately
transforms how network algorithms are developed in the future.

%% file: sections/approach.tex
\section{Our Approach}
\label{sec:approach}

\subsection{Generating Diverse Designs with LLMs}
\label{ssec:sample}

We apply \sysname to the well-known ABR algorithm Pensieve~\cite{mao2017neural},
generating alternative algorithm designs that improve performance with
the assistance of LLMs. Throughout this paper, we use Pensieve as a case study to
demonstrate our methodology. However, we note that \sysname
is not confined to this example; it can be applied to a broader range of network
algorithms, especially RL-based ones.

\begin{figure}[t]
    \centering
    \includegraphics[width=0.8\linewidth]{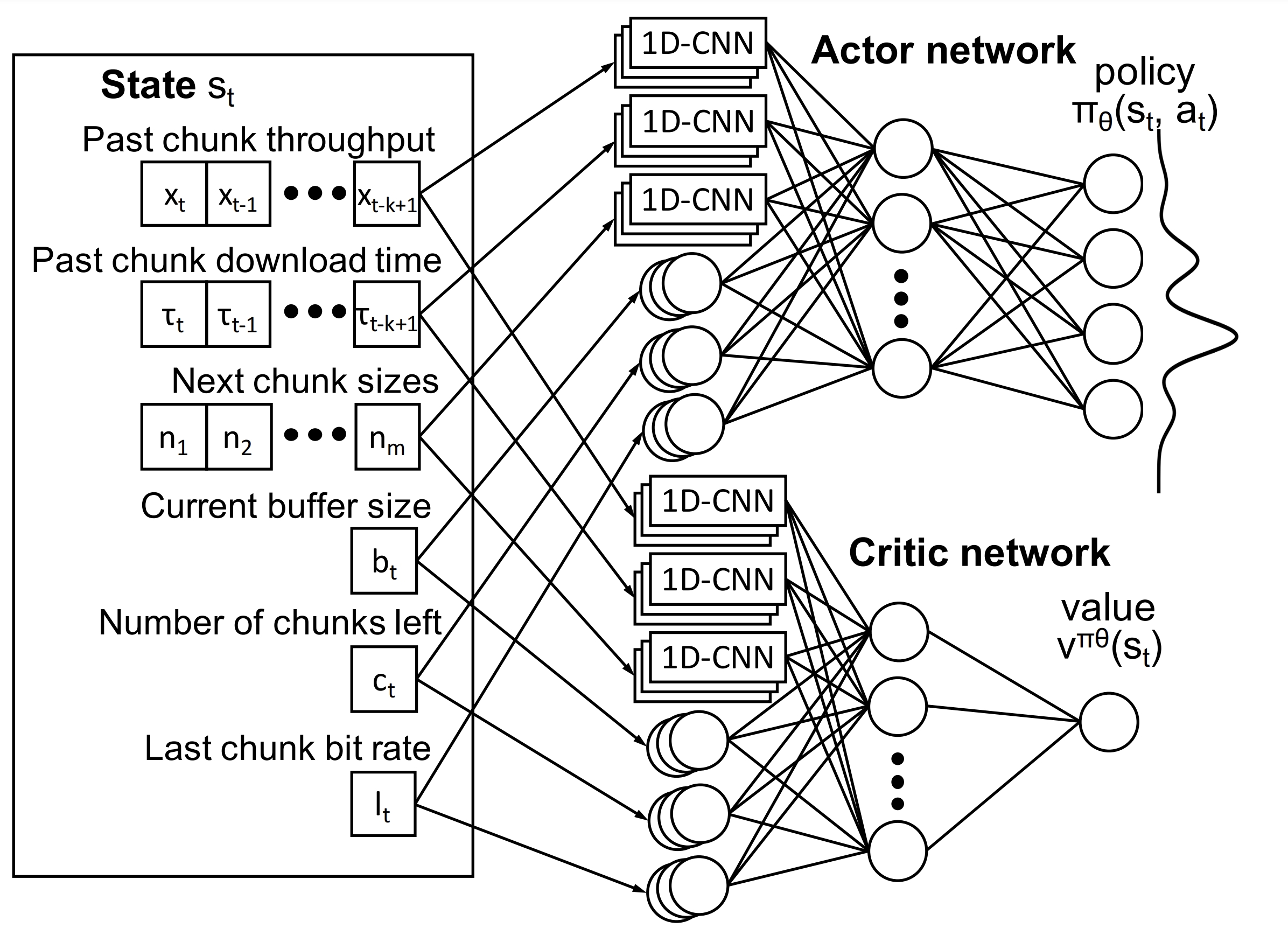}
    \vspace{-5pt}
    \caption{The original algorithm design of Pensieve~\cite{mao2017neural}.}
    \vspace{-10pt}
    \label{fig:pensieve}
\end{figure}

Pensieve leverages an RL method known as the ``actor critic,''
as shown in Figure~\ref{fig:pensieve}.
After streaming each video chunk $t$,
Pensieve constructs a state
$s_t = (\vec{x_t}, \vec{\tau_t}, \vec{n_t}, b_t, c_t, l_t)$
to capture the surrounding network environment.
In this state,
the vectors $\vec{x_t}$, $\vec{\tau_t}$, and $\vec{n_t}$
represent the past network throughput measurements,
the previous download times of video chunks,
and the available sizes of the next chunk at different bitrates, respectively.
The variables $b_t$, $c_t$, and $l_t$ correspond to the current playback buffer
size, the number of remaining chunks in the video, and the last selected bitrate.
Then, the state $s_t$ is input into an actor-critic neural network.
The actor network determines the probability of selecting a particular
bitrate for the next chunk, while the critic network estimates the expected
reward achievable from $s_t$.

It can be seen from Figure~\ref{fig:pensieve} that the Pensieve algorithm
is built around two essential components: the RL state representation and the actor-critic
neural network architecture, both hand-designed and manually implemented through
Python functions (code blocks).
Given these existing code blocks, \sysname first aims to guide LLMs in generating
a wide array of alternative code blocks that potentially encapsulate novel state designs
and neural network architectures.
The main objective is to stimulate diversity and creativity in the algorithm designs
generated by LLMs, thereby increasing the chances of producing high-quality solutions.

Through experimentation, we identified several effective prompting strategies.
First, we instruct LLMs to analyze existing code, generate multiple ideas in natural language,
and then select the best idea before proceeding to code generation.
This method, known as Chain-of-Thought (CoT)~\cite{wei2022chain},
enhances the LLM's reasoning capabilities and leads to more diverse outputs.
Second, we rename the original variables, i.e., the parameters of state
and neural network functions, to more semantically meaningful names.
We further explain their roles both in the prompt and through detailed code comments.
While not strictly necessary, this revision and annotation process helps
LLMs better understand the problem and generate higher-quality solutions.
Lastly, we observe that LLMs sometimes generate state designs with improperly
normalized features, which hinders convergence and degrades performance.
To mitigate this issue, we explicitly request proper normalization in our
state generation prompts. This strategy does not apply to neural network architectures.
The complete set of prompts is released at
\url{https://github.com/hzy46/NADA}.

\subsection{Filtering and Evaluating Designs}
\label{ssec:filter}

The state representations and neural network architectures generated by LLMs
often fall short of expectations.
Given the large number of algorithm designs produced by LLMs,
the main challenge is to efficiently and accurately
evaluate them while identifying promising candidates.
To address this challenge, we develop three filtering strategies
aimed at reducing training and evaluation costs.
The first two strategies serve as pre-checks:
an initial compilation (or execution) check to filter out code with syntax errors,
and an empirical heuristic to remove states with unnormalized features.
The third strategy implements an early stopping mechanism,
using a predictive model to terminate the training of unpromising designs
before they fully complete.
These techniques enable early identification of flawed designs, minimizing
unnecessary computational costs without overlooking promising designs.
Next, we elaborate on the design of the pre-checks and the early stopping mechanism.

The compilation check involves a trial run of the LLM-generated code.
Any code that triggers an exception is immediately excluded from further
consideration.
Following this, a normalization check is applied to the generated states. We
observe that LLMs sometimes use features like chunk sizes in bytes, which
can result in abnormally large values (e.g., over one million for megabytes),
hindering the convergence of the training process.
To eliminate state designs with improperly normalized features, we test the
code with random inputs (``fuzzing''), and check whether any output contains
a feature value exceeding a predefined threshold $T$ (set to 100 in our study).
State designs that fail this test are rejected.
This normalization check is applied only to state generation code,
not the code that defines the neural network architecture.

Once an LLM-generated design passes both the compilation and normalization checks,
\sysname proceeds to train it in a network simulator (or emulator).
However, RL training is computationally expensive,
requiring numerous epochs to reach convergence.
To reduce the cost, we introduce an early stopping model---a binary
classifier---that predicts whether the training trajectory
in the early stages is likely to result in a performant algorithm.
Specifically, this early stopping model utilizes the training rewards
from the first $K$ episodes to learn a 1D-CNN
(one-dimensional convolutional neural network) as the binary classifier.
If the classifier predicts that a particular algorithm design is unlikely
to rank among the top performers, \sysname will early-stop its training.

\begin{table*}[t]
\begin{tabular}{lccccccc}
\toprule
\textbf{Dataset} & \textbf{Train Traces} & \textbf{Train Hours} & \textbf{Test Traces} & \textbf{Test Hours} & \textbf{Throughput} & \textbf{Train Epochs} & \textbf{Test Interval} \\
\midrule
FCC              & 85                    & 10.0                    & 290               & 25.7                & 1.3                 & 40,000                & 500                         \\
Starlink         & 13                    & 0.9                     & 12                & 0.8                 & 1.6                 & 4,000                 & 100                         \\
4G               & 119                   & 10.0                    & 121               & 10.0                & 19.8                & 40,000                & 500                         \\
5G               & 117                   & 10.0                    & 119               & 10.0                & 30.2                & 40,000                & 500                         \\
\bottomrule
\end{tabular}
\vspace{5pt}
\caption{Network traces used in our study. ``Train Traces'' and ``Test Traces'' are the number of traces in the training and testing splits, respectively. ``Train Hours'' and ``Test Hours'' are the total duration of the traces measured in hours. ``Throughput'' represents the average throughput in Mbps. The last two columns show the number of training epochs and the intervals at which model checkpoints are evaluated on the corresponding test sets.}
\label{tab:datasets}
\vspace{-15pt}
\end{table*}

Ideally, the early stopping model would filter out all but the top-performing
designs, such as the top 1\%.
However, labeling only the top 1\% of designs as positive in the training data
leads to poor classification performance due to the significant class imbalance
between the positive class (1\%) and the negative class (99\%).
To address this imbalance, we employ a variant of label
smoothing~\cite{label-smoothing}.
Instead of labeling only the top 1\% as positive,
we expand the positive label to the top 20\%.
This adjustment reduces class skew and enables the early stopping model
to learn more distinguishing characteristics of high-performing designs.
Then, we revert to the original label assignment (top 1\% as positive),
and fine-tune the model's classification threshold on the training set,
i.e., predicting a positive (or negative) label if the model's output score
is above (or below) the threshold.
Since overlooking a performant design has a worse impact than unnecessarily
evaluating a suboptimal design, the threshold is increased to maximize the
true negative rate (unpromising designs correctly early-stopped)
while maintaining a 0\% false negative rate (top-performing designs correctly
preserved).

We compare this model with alternative predictive methods and report
results in \S\ref{sec:early-stopping-eval}. Our results indicate that
the early stopping model can correctly early-stop 87\% of previously unseen
designs without prematurely rejecting any of the top 5 performing designs.

%% file: sections/evaluation.tex
\section{Evaluation}
\label{sec:eval}

\subsection{Experiment Settings}
\label{sec:experiment-settings}

We perform a trace-driven evaluation using the following trace datasets.
Details are presented in Table~\ref{tab:datasets}.

\begin{itemize}[itemsep=1pt,topsep=5pt,leftmargin=*]
    \item \textbf{FCC}: This dataset represents measurements of the U.S. broadband network as recorded by the FCC~\cite{fccMeasuringBroadband}.
    \item \textbf{4G} and \textbf{5G}: We create these two datasets by measuring downlink throughput from 4G and 5G networks in the U.S.
    \item \textbf{Starlink}: We collect throughput traces from a stationary Starlink RV terminal located in the U.S. While Starlink's bandwidth can support high-resolution video streaming during off-peak hours, it decreases significantly during peak
    hours due to shared usage of satellite links. To simulate this condition, we reduce the link capacity in the Starlink traces to one-eighth of its original speed.
\end{itemize}

We adopt the same video streaming configurations as in Pensieve~\cite{mao2017neural},
including the same bitrate levels of 
\{300, 750, 1200, 1850, 2850, 4300\} kbps when evaluating on the FCC and Starlink datasets.
However, since our 4G and 5G datasets exhibit much higher bandwidth,
we elevate the bitrate ladder to \{1850, 2850, 4300, 12000, 24000, 53000\} kbps.
This bitrate ladder follows YouTube's recommended video encoding
settings~\cite{googleYouTubeRecommended}.
The same quality of experience (QoE) function from Pensieve
(``$QoE_{lin}$'') is adopted as the reward.

On each trace dataset, we train both the original Pensieve and the novel designs
generated by \sysname, allowing for algorithm customization in different
network environments.
Two LLMs are tested with \sysname---GPT-3.5 and GPT-4.
To reduce the influence of random noise,
we perform five independent training sessions for each design,
with each session initialized using a different random seed.
During each session, we periodically evaluate model checkpoints
on the \textit{test traces} and calculate the average reward
from the last 10 checkpoints.
The median of these smoothed rewards from the five sessions
is reported as the final ``test score'' (or simply ``score'').
Table~\ref{tab:datasets} lists the number of training epochs and the frequency
of checkpoint testing.

\subsection{Designing States}
\label{sec:designing-states}

\begin{table}[]
\vspace{5pt}
\begin{tabular}{lcrr}
\toprule
    {\sysname}    & \textbf{Total} & \textbf{Compilable} & \textbf{Well Normalized} \\
\midrule
w/ GPT-3.5 & 3,000         & 1,237 (41.2\%)     & 822 (27.4\%)             \\
w/ GPT-4   & 3,000         & 2,059 (68.6\%)     & 1,505 (50.2\%)            \\
\bottomrule
\end{tabular}
\vspace{5pt}
\caption{Number of ABR designs generated by \sysname using GPT-3.5 and GPT-4 that
successfully pass the compilation check and the normalization check.}
\label{tab:state-gpt35-gpt4}
\vspace{-25pt}
\end{table}

\begin{figure*}[htbp]
    \centering
    \includegraphics[width=0.95\linewidth]{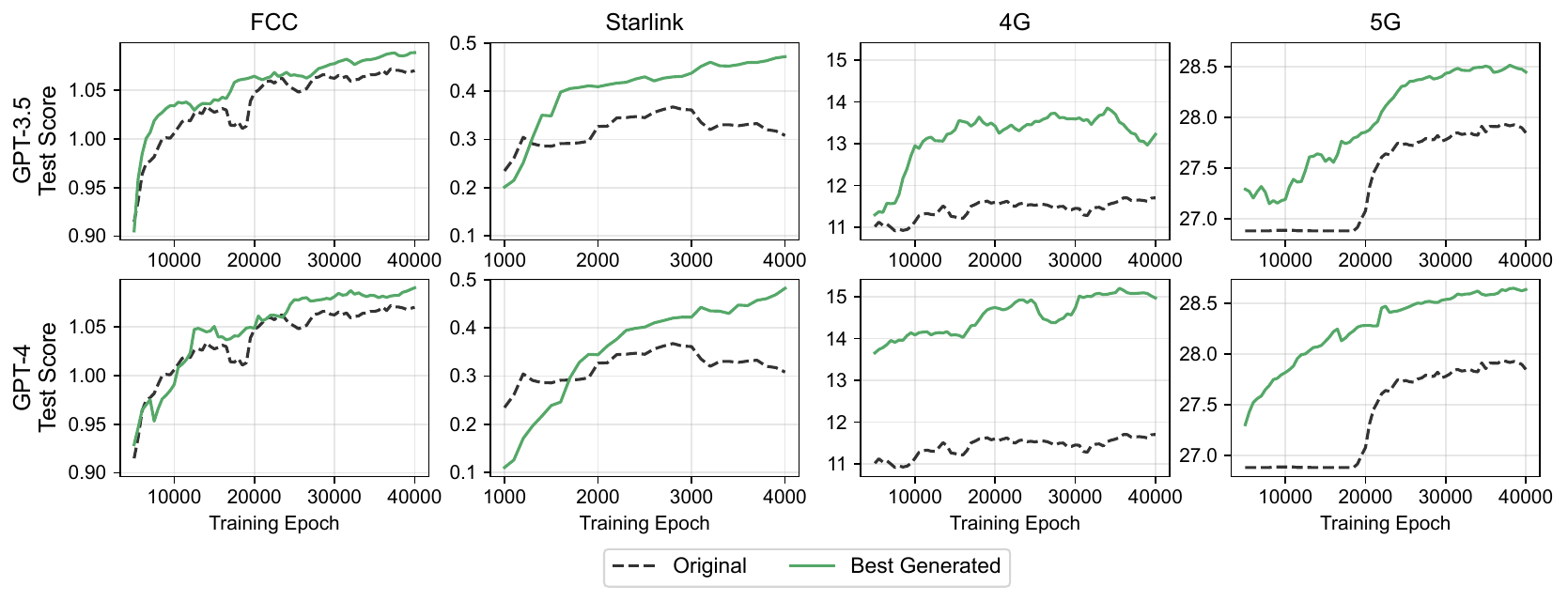}
    \vspace{-10pt}
    \caption{Test performance of the best states generated by \sysname using GPT-3.5
    and GPT-4, compared with the original state design throughout the training process.
    \sysname consistently produces state representations that outperform the
    original design across four network trace sets in simulation.}
    \vspace{-5pt}
    \label{fig:state-result}
\end{figure*}

We run \sysname on GPT-3.5 and GPT-4 to generate 3,000 states each.
The statistics in Table~\ref{tab:state-gpt35-gpt4} show
that 68.6\% of the state functions generated by GPT-4 are ``compilable,''
i.e., they execute without errors, compared with 41.2\% for GPT-3.5.
Meanwhile, 50.2\% of the states produced by GPT-4 contain well-normalized
features, whereas only 27.4\% of those from GPT-3.5 do.
These results highlight GPT-4's superior capability in generating 
correct and desired code blocks.

\begin{table}[t]
\begin{tabular}{llcc}
\toprule
\textbf{Dataset} & \textbf{Method} & \textbf{Score}  & \textbf{Impr.}  \\
\midrule
FCC              & Original         & 1.070           & --             \\
FCC              & w/ GPT-3.5      & 1.089           & 1.7\%           \\
FCC              & w/ GPT-4        & \colorbox{best}{1.090}  & \colorbox{best}{1.9\%}  \\
\midrule
Starlink         & Original         & 0.308           & --             \\
Starlink         & w/ GPT-3.5      & 0.472           & 52.9\%          \\
Starlink         & w/ GPT-4        & \colorbox{best}{0.482}  & \colorbox{best}{56.3\%} \\
\midrule
4G               & Original         & 11.705          & --             \\
4G               & w/ GPT-3.5      & 13.226          & 13.0\%          \\
4G               & w/ GPT-4        & \colorbox{best}{14.973} & \colorbox{best}{27.9\%} \\
\midrule
5G               & Original         & 27.848          & --             \\
5G               & w/ GPT-3.5      & 28.447          & 2.2\%           \\
5G               & w/ GPT-4        & \colorbox{best}{28.636} & \colorbox{best}{2.8\%}  \\
\bottomrule
\end{tabular}
\vspace{5pt}
\caption{Test performance of the best states generated by \sysname using GPT-3.5
and GPT-4 after the training completes. Network traces are replayed in simulation.}
\label{tab:state-design}
\vspace{-15pt}
\end{table}

The alternative state designs proposed by GPT can be non-trivial.
We find that GPT introduces not only basic features, such as
bitrate variance and the exponential moving average of throughput,
but also imports additional Python packages to implement more advanced
functionality. For instance, some states use the linear regression model from the
\texttt{statsmodel} package to predict future throughput.
In another example, the Savitzky-Golay filter~\cite{savitzky1964smoothing}
from the \texttt{scipy} package is applied to analyze buffer size trends
based on historical data. In contrast, the original state representation in
Pensieve does \textit{not} utilize buffer size history in any form.

In Figure~\ref{fig:state-result}, we compare the best states generated by
\sysname using GPT-3.5 and GPT-4 against the original state design.
The test scores (as defined in \S\ref{sec:experiment-settings})
are plotted throughout the training sessions for each network trace set.
Table~\ref{tab:state-design} provides a summary of the final test scores
after the maximum number of training epochs.
These results show that \sysname, when applied with both GPT-3.5 and GPT-4,
consistently generates state representations that outperform the original design,
with GPT-4 demonstrating a more significant overall improvement,
especially on the Starlink traces.

In addition, we conduct emulation experiments using the dash.js framework 
to stream video in a real web browser over Mahimahi~\cite{mahimahi}.
The results for the Starlink, 4G and 5G traces are shown in
Table~\ref{tab:emulation} (we did not evaluate on FCC as the simulation improvements
were already statistically insignificant).
Despite discrepancies in the emulation and simulation results, the optimal states
generated by \sysname continue to outperform the original design.
In Section~\ref{sec:insight}, we elaborate on the best states generated
for each network scenario and provide insights into their design.

\begin{table}[]
\begin{tabular}{llcc}
\toprule
\textbf{Dataset} & \textbf{Method} & \textbf{Score}           & \textbf{Impr.}                             \\ \midrule
Starlink         & Original         & \textminus 0.0482                   & --                \\
Starlink         & w/ GPT-3.5      & \colorbox{best}{0.0899}  & \colorbox{best}{286.5\%}                         \\
Starlink         & w/ GPT-4        & 0.0759                   & 257.5\%                 \\ \midrule
4G               & Original         & 4.976                    & --                 \\
4G               & w/ GPT-3.5      &  8.010                   & 61.0\% \\               
4G               & w/ GPT-4        & \colorbox{best}{9.233}   & \colorbox{best}{85.6\%}                         \\ \midrule
5G               & Original         & 17.26                    & --                 \\
5G               & w/ GPT-3.5      & 17.43                    & 1.0\%                 \\
5G               & w/ GPT-4        & \colorbox{best}{21.55}   & \colorbox{best}{24.9\%}                         \\ \bottomrule
\end{tabular}
\vspace{5pt}
\caption{Emulation results of the best generated states.}
\label{tab:emulation}
\vspace{-25pt}
\end{table}

\subsection{Designing Neural Networks}
\label{sec:designing-network-arch}

\begin{figure*}[t]
    \centering
    \includegraphics[width=0.95\linewidth]{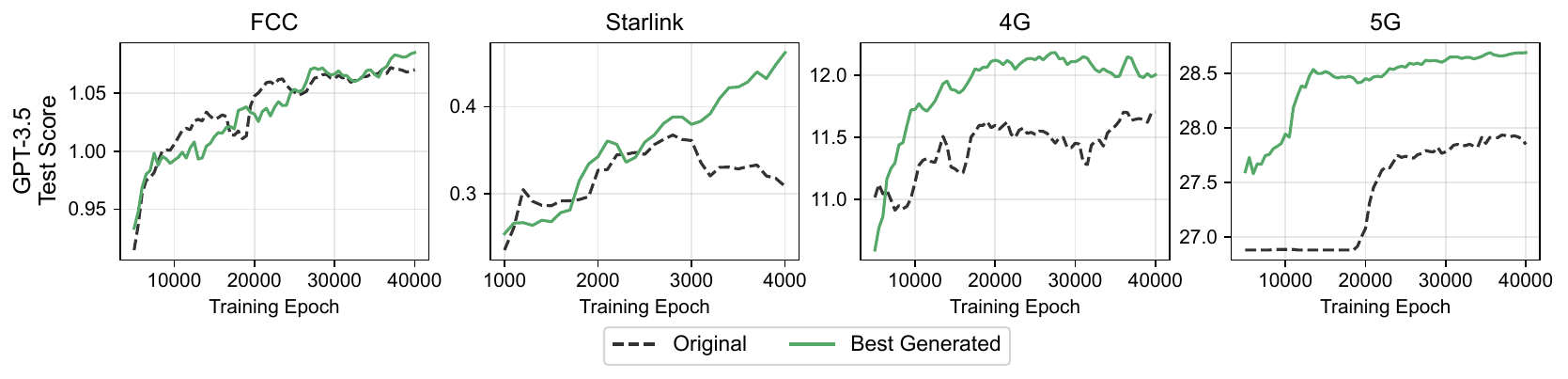}
    \vspace{-10pt}
    \caption{Test performance of the best generated neural network architectures
    vs. the original in simulation.}
    \label{fig:network-arch-result}
\end{figure*}

Due to budget constraints, our investigation into the neural network architecture
is restricted to GPT-3.5. We run \sysname on GPT-3.5 to generate 3,000
alternative architectures and apply the compilation check to filter out invalid
designs (the normalization check is not applicable here).
Among the generated neural networks, 760 architectures pass
the compilation check.
Figure~\ref{fig:network-arch-result} compares the most
effective architectures with the original design.
Notably, more pronounced improvements are observed on the
Starlink, 4G, and 5G traces,
whereas the improvement on FCC is not statistically
significant.
Overall, we find that modifying the neural network architecture
tends to yield smaller gains than revising the state.
The emulation results are omitted here.

Furthermore, we explore the performance improvements by 
combining novel states with newly generated neural network architectures.
Specifically, we select the top 30 states and the top 30
neural networks generated by GPT-3.5, creating 900 unique combinations.
Each state-architecture combination is trained five times,
and the best results are presented in Table~\ref{tab:combination}.
We find that this combination leads to consistent improvements,
with gains up to 61.1\% on the Starlink traces.
Nevertheless, the combined improvements are relatively modest compared with
the individual gains from updating states or neural networks alone.

\begin{table}[]
\begin{tabular}{lccc}
\toprule
\textbf{Dataset} & \textbf{State} & \textbf{Neural Net} & \textbf{Combined} \\
\midrule 
FCC              & 1.7\%          & 1.4\%                 & \colorbox{best}{2.2\%}                \\
Starlink         & 52.9\%         & 50.0\%                & \colorbox{best}{61.1\%}               \\
4G               & 13.0\%         & 2.6\%                 & \colorbox{best}{16.5\%}                         \\
5G               & 2.2\%          & 3.0\%                 & \colorbox{best}{3.1\%}                \\
\bottomrule 
\end{tabular}
\vspace{5pt}
\caption{Results of combining the states and neural networks
generated by \sysname with GPT-3.5.}
\label{tab:combination}
\vspace{-15pt}
\end{table}

\subsection{Early Stopping Mechanism}
\label{sec:early-stopping-eval}

\begin{figure*}[htbp]
    \centering
    \includegraphics[width=0.9\linewidth]{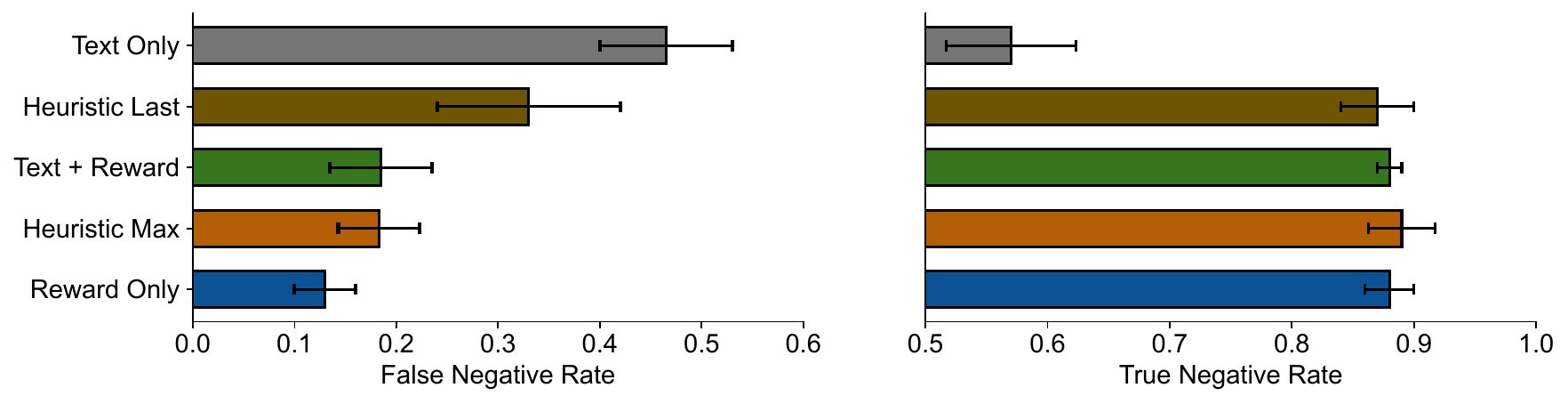}
    \caption{Comparison between different early stopping classifiers. False and true negative rates are defined in \S\ref{sec:early-stopping-eval}.}
    \label{fig:early-stop-evals}
\end{figure*}

In this section, we introduce and assess five candidate mechanisms for
early stopping during training.
We consider alternative designs, including novel states or neural network
architectures, that fall within the top 1\% of training rewards as candidates worth full training.
These top designs are labeled as positive, while the remaining ones are labeled negative.
The methods tested are as follows:
(1) ``Reward Only'': Utilizing the first 10k training rewards to learn a 1D-CNN classifier;
(2) ``Text Only'': Embedding the code using OpenAI's \texttt{text-embedding-ada-002}
model as input to the trained classifier;
(3) ``Text + Reward'': Using the previous two features as inputs to the classifier;
(4) ``Heuristic Max'': Early stopping based on the maximum reward in the first 10k epochs;
(5) ``Heuristic Last'': Early stopping based on the reward in the final epoch.

We first collect 2000 algorithm designs along with their corresponding training metrics,including ground-truth labels, and conduct a five-fold cross validation. In each fold, 20\% of the designs, or 400 samples, are used for training. We report two metrics across all
validation folds and network environments:
the false negative rate---fraction of top-performing designs incorrectly reject,
and the true negative rate---fraction of suboptimal designs correctly stopped early.

On the samples for testing, Figure~\ref{fig:early-stop-evals} shows that
``Reward Only'' offers the best trade-off between early stopping errors (left panel) and resource savings (right panel). Specifically, ``Reward Only'' successfully terminates 87\% of suboptimal designs with an incorrect rejection rate of only 12\% on average. We also manually confirmed that the top five algorithms are never missed. This translates to computational savings on the order of hundreds of millions of training epochs.

%% file: sections/insight.tex
\section{Insights from Generated Designs}
\label{sec:insight}

In this section, we describe the best designs generated by \sysname using GPT-3.5 and GPT-4 for
each network environment, with a focus on the innovative state designs and the
key insights gained from them. We then provide a short
summary of the changes introduced in the neural network
architectures before concluding this section.

\para{FCC:} On the FCC traces, we observe that the optimal states generated
by \sysname using both GPT-3.5 and GPT-4 involve modifying the normalization strategy
for certain features.
While the original normalization range is $[0, 1]$,
the optimal states remap these features to $[-1, 1]$.

\para{Starlink:} \sysname with GPT-3.5 exploits the smaller size of the Starlink
dataset and removes two variables from the state representation:
the download times of previous video chunks,
and the size options for the next chunk.
This approach seems to reduce overfitting and showcases \sysname's ability to 
autonomously customize network algorithms based on environmental complexity
(reflected as the number of traces in our study),
but we did not empirically verify this claim.
In comparison, GPT-4 employs more aggressive normalization
with an increased normalizing factor
and smooths the throughput and download times.

\para{4G:} On the 4G traces, the original state tends to favor lower bitrates,
leading to lower rewards.
Consequently, the optimal states generated by \sysname introduce new features
that promote higher bitrate selection
when the video playback is sufficiently buffered.
GPT-3.5, for instance, applies a linear regression model to predict the download
time of future chunks and incorporates the trends of throughput and download time
into the state.
GPT-4 introduces the historical trend of playback buffer size,
signaling the model to increase the bitrate as the buffer grows.
In contrast, the original state design does not take buffer size history
into account at all.

\para{5G:} The best states for the 5G traces are similar to those for 4G.
GPT-3.5 introduces a predicted throughput feature, while GPT-4 adds the
buffer size difference between adjacent time steps.
These enhancements allow the model to make more informed bitrate decisions
and achieve higher rewards.

\para{Summary:} LLMs have suggested several principles for designing the states
of RL-based ABR algorithms.
First, selecting an appropriate normalization strategy (with a different normalization range or normalizing factor)
can enhance model performance.
Second, removing unnecessary state features \textit{might}
reduce overfitting particularly in simpler target environments.
Third, even though a 1D-CNN can implicitly capture past throughputs
and download times, explicitly summarizing their trends or predicting
future values as additional features may still provide benefits.
We hypothesize that this extra layer of feature engineering
helps preserve important signals amid noisy data.
Finally, and perhaps most intriguingly,
Pensieve has overlooked the relevance of buffer size history in ABR.
In contrast, \sysname reveals that incorporating features like buffer
size trends or differences (between adjacent time steps)
leads to noticeable improvements.

\vspace{0.5\baselineskip}
Next, we briefly summarize the key changes introduced by the best generated neural network architectures.
For the FCC traces, the number of hidden neurons in the fully connected network is increased to 256, and the activation function is switched to Leaky ReLU.
For Starlink, an RNN is used in place of a 1D-CNN, while in 4G,
an LSTM is used instead.
For the 5G dataset, the actor and critic networks share the hidden layer
but retain separate output layers.
Complete results are available at \url{https://github.com/hzy46/NADA}.

%% file: sections/discussion.tex
\section{Discussion}
\label{sec:discussion}

Through our exploration of applying \sysname to ABR algorithms, we have learned
the following lessons. First, directly applying LLMs to optimize large,
complex programs proves challenging, while optimizing 
individual functions (e.g., states or neural networks) enables
more manageable and targeted improvements. 
Second, LLMs can generate creative design alternatives,
but not all suggestions are useful. Therefore, it is essential to develop
efficient filtering mechanisms to quickly assess the quality of LLM-generated designs.

Our work demonstrates the potential of LLMs in designing network algorithms,
and we identify several promising future directions:
(1) We plan to extend the case study from ABR to other network algorithms,
such as congestion control.
(2) While our current focus is on enhancing RL-based algorithms,
we believe \sysname can be adapted to generate other types of
network algorithms, although different filtering techniques may be required.
(3) LLMs have demonstrated the ability to propose creative algorithmic
modifications, but their proposals lack completeness and rigor.
Moving forward, we aim to integrate LLMs with program synthesis
or neural architecture search (NAS) to systematically explore the design space.
(4) Lastly, we intend to refine our prompting strategies and more effectively
harness the reasoning capabilities of LLMs to reduce the number of initial
candidates, while still maintaining diversity and quality. This will improve
the overall efficiency of our framework.

%% file: sections/conclusion.tex
\section{Conclusion}
\label{sec:conclusion}

In this paper, we presented \sysname, a framework that leverages LLMs
to develop novel network algorithms tailored to diverse network environments.
Using ABR as a case study, we demonstrated that \sysname
effectively generated novel RL state designs and neural network architectures
that consistently outperformed
the original ABR design in different network environments.
In future work, we plan to extend our framework beyond ABR to other network algorithms.

%% file: hotnets24-nada.bbl

\begin{thebibliography}{21}


\ifx \showCODEN    \undefined \def \showCODEN     #1{\unskip}     \fi
\ifx \showDOI      \undefined \def \showDOI       #1{#1}\fi
\ifx \showISBNx    \undefined \def \showISBNx     #1{\unskip}     \fi
\ifx \showISBNxiii \undefined \def \showISBNxiii  #1{\unskip}     \fi
\ifx \showISSN     \undefined \def \showISSN      #1{\unskip}     \fi
\ifx \showLCCN     \undefined \def \showLCCN      #1{\unskip}     \fi
\ifx \shownote     \undefined \def \shownote      #1{#1}          \fi
\ifx \showarticletitle \undefined \def \showarticletitle #1{#1}   \fi
\ifx \showURL      \undefined \def \showURL       {\relax}        \fi
\providecommand\bibfield[2]{#2}
\providecommand\bibinfo[2]{#2}
\providecommand\natexlab[1]{#1}
\providecommand\showeprint[2][]{arXiv:#2}

\bibitem[Chen et~al\mbox{.}(2021)]%
        {chen2021evaluating}
\bibfield{author}{\bibinfo{person}{Mark Chen}, \bibinfo{person}{Jerry Tworek}, \bibinfo{person}{Heewoo Jun}, \bibinfo{person}{Qiming Yuan}, \bibinfo{person}{Henrique Ponde de~Oliveira Pinto}, \bibinfo{person}{Jared Kaplan}, \bibinfo{person}{Harri Edwards}, \bibinfo{person}{Yuri Burda}, \bibinfo{person}{Nicholas Joseph}, \bibinfo{person}{Greg Brockman}, {et~al\mbox{.}}} \bibinfo{year}{2021}\natexlab{}.
\newblock \showarticletitle{Evaluating large language models trained on code}.
\newblock \bibinfo{journal}{\emph{arXiv preprint arXiv:2107.03374}} (\bibinfo{year}{2021}).
\newblock


\bibitem[FCC(2024)]%
        {fccMeasuringBroadband}
\bibfield{author}{\bibinfo{person}{FCC}.} \bibinfo{year}{2024}\natexlab{}.
\newblock \bibinfo{title}{{M}easuring {B}roadband {A}merica}.
\newblock \bibinfo{howpublished}{\url{https://www.fcc.gov/general/measuring-broadband-america}}.
\newblock
\newblock
\shownote{[Accessed 10-03-2024]}.


\bibitem[Google(2024)]%
        {googleYouTubeRecommended}
\bibfield{author}{\bibinfo{person}{Google}.} \bibinfo{year}{2024}\natexlab{}.
\newblock \bibinfo{title}{{Y}ou{T}ube recommended upload encoding settings}.
\newblock \bibinfo{howpublished}{\url{https://support.google.com/youtube/answer/1722171?hl=en}}.
\newblock
\newblock
\shownote{[Accessed 10-03-2024]}.


\bibitem[Hong et~al\mbox{.}(2023)]%
        {hong2023metagpt}
\bibfield{author}{\bibinfo{person}{Sirui Hong}, \bibinfo{person}{Xiawu Zheng}, \bibinfo{person}{Jonathan Chen}, \bibinfo{person}{Yuheng Cheng}, \bibinfo{person}{Jinlin Wang}, \bibinfo{person}{Ceyao Zhang}, \bibinfo{person}{Zili Wang}, \bibinfo{person}{Steven Ka~Shing Yau}, \bibinfo{person}{Zijuan Lin}, \bibinfo{person}{Liyang Zhou}, {et~al\mbox{.}}} \bibinfo{year}{2023}\natexlab{}.
\newblock \showarticletitle{{MetaGPT}: Meta programming for a multi-agent collaborative framework}.
\newblock \bibinfo{journal}{\emph{arXiv preprint arXiv:2308.00352}} (\bibinfo{year}{2023}).
\newblock


\bibitem[Jay et~al\mbox{.}(2019)]%
        {jay2019deep}
\bibfield{author}{\bibinfo{person}{Nathan Jay}, \bibinfo{person}{Noga Rotman}, \bibinfo{person}{Brighten Godfrey}, \bibinfo{person}{Michael Schapira}, {and} \bibinfo{person}{Aviv Tamar}.} \bibinfo{year}{2019}\natexlab{}.
\newblock \showarticletitle{A deep reinforcement learning perspective on internet congestion control}. In \bibinfo{booktitle}{\emph{International Conference on Machine Learning}}.
\newblock


\bibitem[Kumar et~al\mbox{.}(2016)]%
        {kumar2016adaptive}
\bibfield{author}{\bibinfo{person}{Dhananjay Kumar}, \bibinfo{person}{S. Aishwarya}, \bibinfo{person}{A. Srinivasan}, {and} \bibinfo{person}{L.~Arun Raj}.} \bibinfo{year}{2016}\natexlab{}.
\newblock \showarticletitle{Adaptive video streaming over HTTP using stochastic bitrate prediction in 4G wireless networks}. In \bibinfo{booktitle}{\emph{2016 ITU Kaleidoscope: ICTs for a Sustainable World (ITU WT)}}.
\newblock


\bibitem[Mao et~al\mbox{.}(2019)]%
        {mao2019park}
\bibfield{author}{\bibinfo{person}{Hongzi Mao}, \bibinfo{person}{Parimarjan Negi}, \bibinfo{person}{Akshay Narayan}, \bibinfo{person}{Hanrui Wang}, \bibinfo{person}{Jiacheng Yang}, \bibinfo{person}{Haonan Wang}, \bibinfo{person}{Ryan Marcus}, \bibinfo{person}{Mehrdad Khani~Shirkoohi}, \bibinfo{person}{Songtao He}, \bibinfo{person}{Vikram Nathan}, {et~al\mbox{.}}} \bibinfo{year}{2019}\natexlab{}.
\newblock \showarticletitle{Park: An open platform for learning-augmented computer systems}.
\newblock \bibinfo{journal}{\emph{Advances in Neural Information Processing Systems}} (\bibinfo{year}{2019}).
\newblock


\bibitem[Mao et~al\mbox{.}(2017)]%
        {mao2017neural}
\bibfield{author}{\bibinfo{person}{Hongzi Mao}, \bibinfo{person}{Ravi Netravali}, {and} \bibinfo{person}{Mohammad Alizadeh}.} \bibinfo{year}{2017}\natexlab{}.
\newblock \showarticletitle{Neural Adaptive Video Streaming with Pensieve}. In \bibinfo{booktitle}{\emph{Proceedings of the conference of the ACM special interest group on data communication}}.
\newblock


\bibitem[M{\"u}ller et~al\mbox{.}(2019)]%
        {label-smoothing}
\bibfield{author}{\bibinfo{person}{Rafael M{\"u}ller}, \bibinfo{person}{Simon Kornblith}, {and} \bibinfo{person}{Geoffrey~E. Hinton}.} \bibinfo{year}{2019}\natexlab{}.
\newblock \showarticletitle{When does label smoothing help?}
\newblock \bibinfo{journal}{\emph{Advances in Neural Information Processing Systems}} (\bibinfo{year}{2019}).
\newblock


\bibitem[Netravali et~al\mbox{.}(2015)]%
        {mahimahi}
\bibfield{author}{\bibinfo{person}{Ravi Netravali}, \bibinfo{person}{Anirudh Sivaraman}, \bibinfo{person}{Somak Das}, \bibinfo{person}{Ameesh Goyal}, \bibinfo{person}{Keith Winstein}, \bibinfo{person}{James Mickens}, {and} \bibinfo{person}{Hari Balakrishnan}.} \bibinfo{year}{2015}\natexlab{}.
\newblock \showarticletitle{Mahimahi: accurate record-and-replay for {HTTP}}. In \bibinfo{booktitle}{\emph{2015 USENIX Annual Technical Conference (USENIX ATC '15)}}.
\newblock


\bibitem[Ramadan et~al\mbox{.}(2021)]%
        {ramadan2021case}
\bibfield{author}{\bibinfo{person}{Eman Ramadan}, \bibinfo{person}{Arvind Narayanan}, \bibinfo{person}{Udhaya~Kumar Dayalan}, \bibinfo{person}{Rostand~AK Fezeu}, \bibinfo{person}{Feng Qian}, {and} \bibinfo{person}{Zhi-Li Zhang}.} \bibinfo{year}{2021}\natexlab{}.
\newblock \showarticletitle{Case for 5G-aware video streaming applications}. In \bibinfo{booktitle}{\emph{Proceedings of the 1st workshop on {5G} measurements, modeling, and use cases}}.
\newblock


\bibitem[Savitzky and Golay(1964)]%
        {savitzky1964smoothing}
\bibfield{author}{\bibinfo{person}{Abraham Savitzky} {and} \bibinfo{person}{Marcel J.~E. Golay}.} \bibinfo{year}{1964}\natexlab{}.
\newblock \showarticletitle{Smoothing and differentiation of data by simplified least squares procedures.}
\newblock \bibinfo{journal}{\emph{Analytical chemistry}} (\bibinfo{year}{1964}).
\newblock


\bibitem[Tran et~al\mbox{.}(2020)]%
        {tran2020bitrate}
\bibfield{author}{\bibinfo{person}{Anh-Tien Tran}, \bibinfo{person}{Nhu-Ngoc Dao}, {and} \bibinfo{person}{Sungrae Cho}.} \bibinfo{year}{2020}\natexlab{}.
\newblock \showarticletitle{Bitrate adaptation for video streaming services in edge caching systems}.
\newblock \bibinfo{journal}{\emph{IEEE Access}} (\bibinfo{year}{2020}).
\newblock


\bibitem[Tuysuz and Aydin(2020)]%
        {tuysuz2020qoe}
\bibfield{author}{\bibinfo{person}{Mehmet~Fatih Tuysuz} {and} \bibinfo{person}{Mehmet~Emin Aydin}.} \bibinfo{year}{2020}\natexlab{}.
\newblock \showarticletitle{QoE-based mobility-aware collaborative video streaming on the edge of 5G}.
\newblock \bibinfo{journal}{\emph{IEEE Transactions on Industrial Informatics}} (\bibinfo{year}{2020}).
\newblock


\bibitem[Wei et~al\mbox{.}(2022)]%
        {wei2022chain}
\bibfield{author}{\bibinfo{person}{Jason Wei}, \bibinfo{person}{Xuezhi Wang}, \bibinfo{person}{Dale Schuurmans}, \bibinfo{person}{Maarten Bosma}, \bibinfo{person}{Fei Xia}, \bibinfo{person}{Ed Chi}, \bibinfo{person}{Quoc~V. Le}, \bibinfo{person}{Denny Zhou}, {et~al\mbox{.}}} \bibinfo{year}{2022}\natexlab{}.
\newblock \showarticletitle{Chain-of-thought prompting elicits reasoning in large language models}.
\newblock \bibinfo{journal}{\emph{Advances in Neural Information Processing Systems}} (\bibinfo{year}{2022}).
\newblock


\bibitem[Xia et~al\mbox{.}(2022)]%
        {xia2022genet}
\bibfield{author}{\bibinfo{person}{Zhengxu Xia}, \bibinfo{person}{Yajie Zhou}, \bibinfo{person}{Francis~Y. Yan}, {and} \bibinfo{person}{Junchen Jiang}.} \bibinfo{year}{2022}\natexlab{}.
\newblock \showarticletitle{Genet: automatic curriculum generation for learning adaptation in networking}. In \bibinfo{booktitle}{\emph{Proceedings of the ACM SIGCOMM 2022 Conference}}.
\newblock


\bibitem[Yan et~al\mbox{.}(2020)]%
        {puffer}
\bibfield{author}{\bibinfo{person}{Francis~Y. Yan}, \bibinfo{person}{Hudson Ayers}, \bibinfo{person}{Chenzhi Zhu}, \bibinfo{person}{Sadjad Fouladi}, \bibinfo{person}{James Hong}, \bibinfo{person}{Keyi Zhang}, \bibinfo{person}{Philip Levis}, {and} \bibinfo{person}{Keith Winstein}.} \bibinfo{year}{2020}\natexlab{}.
\newblock \showarticletitle{{Learning \textit{in situ}: a randomized experiment in video streaming}}. In \bibinfo{booktitle}{\emph{USENIX Symposium on Networked Systems Design and Implementation (NSDI '20)}}.
\newblock


\bibitem[Yan et~al\mbox{.}(2018)]%
        {pantheon}
\bibfield{author}{\bibinfo{person}{Francis~Y. Yan}, \bibinfo{person}{Jestin Ma}, \bibinfo{person}{Greg~D. Hill}, \bibinfo{person}{Deepti Raghavan}, \bibinfo{person}{Riad~S. Wahby}, \bibinfo{person}{Philip Levis}, {and} \bibinfo{person}{Keith Winstein}.} \bibinfo{year}{2018}\natexlab{}.
\newblock \showarticletitle{{Pantheon: the training ground for Internet congestion-control research}}. In \bibinfo{booktitle}{\emph{USENIX Annual Technical Conference (ATC '18)}}.
\newblock


\bibitem[Yin et~al\mbox{.}(2015)]%
        {yin2015control}
\bibfield{author}{\bibinfo{person}{Xiaoqi Yin}, \bibinfo{person}{Abhishek Jindal}, \bibinfo{person}{Vyas Sekar}, {and} \bibinfo{person}{Bruno Sinopoli}.} \bibinfo{year}{2015}\natexlab{}.
\newblock \showarticletitle{A control-theoretic approach for dynamic adaptive video streaming over HTTP}. In \bibinfo{booktitle}{\emph{Proceedings of the ACM SIGCOMM 2015 Conference}}.
\newblock


\bibitem[Zhao and Pan(2023)]%
        {zhao2023qoe}
\bibfield{author}{\bibinfo{person}{Jinwei Zhao} {and} \bibinfo{person}{Jianping Pan}.} \bibinfo{year}{2023}\natexlab{}.
\newblock \showarticletitle{QoE-driven joint decision-making for multipath adaptive video streaming}. In \bibinfo{booktitle}{\emph{2023 IEEE Global Communications Conference (GLOBECOM)}}.
\newblock


\bibitem[Zhou et~al\mbox{.}(2022)]%
        {zhou2022docprompting}
\bibfield{author}{\bibinfo{person}{Shuyan Zhou}, \bibinfo{person}{Uri Alon}, \bibinfo{person}{Frank~F. Xu}, \bibinfo{person}{Zhiruo Wang}, \bibinfo{person}{Zhengbao Jiang}, {and} \bibinfo{person}{Graham Neubig}.} \bibinfo{year}{2022}\natexlab{}.
\newblock \showarticletitle{{DocPrompting}: Generating code by retrieving the docs}.
\newblock \bibinfo{journal}{\emph{arXiv preprint arXiv:2207.05987}} (\bibinfo{year}{2022}).
\newblock


\end{thebibliography}
